\documentclass[a4paper,fleqn,usenatbib]{mnras}
\usepackage{longtable}
\usepackage{graphicx}	% Including figure files
\title[The glitch-induced radio emission changes of PSR B2035+36]{The Spin-down State Change and Mode Change Associated with Glitch Activity of PSR B2035$+$36}
\author[F.F. Kou et al]{
F. F. Kou,$^{1,2}$ J. P. Yuan,$^{1,3}$\thanks{E-mail: yuanjp@xao.ac.cn}  N. Wang, $^{1,3}$ \thanks{na.wang@xao.ac.cn} W. M. Yan,$^{1}$ S. J. Dang$^{1,2}$
\\
$^{1}$Xinjiang Astronomical Observatory, CAS, 150 Science 1-Street, Urumqi, Xinjiang 830011, China\\
$^{2}$University of Chinese Academic of Science, 19A Yuquan Road, Beijing 100049, China\\
$^{3}$Key Laboratory of Radio Astronomy, Chinese Academy of Sciences, Nanjing 210008, China}

\date{Accepted XXX. Received YYY; in original form ZZZ}

\pubyear{2017}

\begin{document}
\label{firstpage}
\pagerange{\pageref{firstpage}--\pageref{lastpage}}
\maketitle

\begin{abstract}
We presented timing results of PSR B2035$+$36 using $\sim 9$-yr observations with the Nanshan 25-m radio telescope. PSR B2035$+$36 was reported to exhibit significant changes in pulse profile correlated with spin-down state variations. We found that the pulsar underwent a glitch with a jump in the frequency of $\Delta{\nu}\sim12.4(5)\, \rm nHz$ around MJD 52950. Unusually, the spin-down rate increased persistently over $800$ days after the glitch, and the average spin-down rate of post-glitch was about $9.6\%$ larger than pre-glitch. Accompanied with the glitch activity, pulse profile became narrower. The pulsar began to switch between two emission modes after the glitch, with pulse width ($W_{\ 50 \rm mean}$) of $8.5(7)^{\circ}$ and $3.7(3)^{\circ}$, respectively. Besides that, the relatively narrow pulse profile gradually became dominant. All of the observations indicate that there should be connection between magnetospheric behavior and glitch activity. We discuss one possibility of magnetosphere fluctuation triggered by glitch event.

\end{abstract}


%%%%%%%%%%%%%%%%%%%%%%%%%%%%%%%%%%%%%%%%%%%%%%%%%%

%%%%%%%%%%%%%%%%% BODY OF PAPER %%%%%%%%%%%%%%%%%%

\section{Introduction}
Pulsars are rapidly rotating and highly magnetized neutron stars. They are taken as the most stable ``clocks'' in the universe because their stable rotation could be used to probe the interstellar medium and gravitational wave \citep{2013_LeeKJ_gravity}. However, pulsar timing observations are also dominated by two categories of irregularities, ``glitch'' and timing noise. Pulsar glitch is a discontinuous and abrupt change in rotation speed, characterized by a sudden increase in spin frequency, and often followed by a recovery process \citep{1969_Baym_glitch}. It is generally regarded as an internal origin, caused either by the crust quake \citep{1991_Ruderman_star_quake,1998_Ruderman_star_quake} or by a sudden transfer of angular momentum from the crustal superfluid to the rest part of the star \citep{1984_Alpar_vortex_creep}. Timing noise is a fairly continuous erratic behavior. The origin of timing noise is poorly understood. \citet{2010_Lyne_magnetosphere} found that timing noises of a few pulsars are quasi-periodic. Remarkably, pulse profiles and spin-down rates of six pulsars also switch between different states. These observations indicate that timing noise may be linked to the magnetospheric change.

A more unusual aspect of the pulsar rotation is that the timing behaviors of a few pulsars coincided with the change of their pulse profiles.  PSR J1119$-$6127 was the first pulsar to show a change in pulse profile following a glitch event \citep{2011_Weltevrede_1119}. Besides that, observations of PSRs J0742$-$2822 and J2021$+$4026 also showed changes in spin-down state and emission state associated with glitch activities \citep{2013_keith-0742,2017_Zhao_2021}. All of these observations indicate that the magnetospheric fluctuation should be linked to glitch activity. PSR B2035$+$36 was also pronounced that a change in the integrated pulse profile was associated with an increase in spin-down rate \citep{2010_Lyne_magnetosphere}. In addition to that, we found that these significant changes in spin-down rates and pulse profiles were associated with a glitch activity.

PSR B2035$+$36 (J2037$+$3621) was discovered by searching for low-luminosity pulsars \citep{1985_dewey}. It is an isolated radio pulsar with period of $0.6187 \, \rm s$ and period derivative of $4.5024\times10^{-15} \, \rm s/s$ \citep{2004_Hobbs_pulsars}. These parameters imply a middle characteristic age of $\tau_{\rm c}\sim2.18\, \rm Myr$ and dipole magnetic field of $B_{\rm s}\sim 1.69\times 10^{12} \, \rm G$ by assuming magnetic dipole braking. In this paper, we report one glitch event, the associated spin-down state change and mode change in PSR B2035$+$36.
%Observation methods are introduced in Section \ref{obs}. The timing results and mode change analyses are presented in Section \ref{res}. Discussions and Conclusions are listed in Section \ref{dis} and Section \ref{con}, respectively.
\section{Observation Methods}
\label{obs}
Pulsar timing observations of Xinjiang Astronomical Observatory (XAO) are carried out by the $25$-m telescope at Nanshan. The receiver has a bandwidth of $320 \, \rm MHz$ centered at $1540 \, \rm MHz$. An analogue filterbank (AFB) with $128 \times 2.5 \, \rm MHz$ sub-channels was used to obtain data before $2010$. After January 2010, the observation data was obtained by a digital filterbank (DFB) with $1024\times0.5 \, \rm MHz$ sub-channels. PSR B2035$+$36 has been generally observed three times per month. The integrate times were $4\sim16$ minutes .

Off-line data were de-dispersed and summed in time and frequency to produce total intensity profiles, and the total intensity profiles were cross-correlated with standard profiles to determine local pulse time-of-arrivals (ToAs). Local ToAs were corrected to the Solar-system using the standard timing program \textsc{TEMPO2} \footnote{http://www.atnf.csiro.au/research/pulsar/tempo2/} \citep{2006_Edwards_Tempo2,2006_Hobbs_Tempo2} with Jet Propulsion Laboratories planetary ephemeris DE $421$ \citep{2009_folkner_DE421}. ToAs were weighted by inverse square of their uncertainties. 

The expression of the pulse phase $\phi$ of the standard timing model is:
\begin{equation}
\phi(t)=\phi_{0}+\nu(t-t_{0})+\frac{1}{2}\dot{\nu}(t-t_{0})^{2}+\frac{1}{6}\ddot{\nu}(t-t_{0})^{3} \, ,
\label{timing}
\end{equation}
where $\phi_{0}$ is the phase at time $t_{0}$, $\nu$, $\dot{\nu}$ and $\ddot{\nu}$ are pulse frequency and its derivatives, respectively.

Pulsar glitch is described by a combination of step changes of $\nu$ and $\dot{\nu}$:
\begin{equation}
\nu(t) = \nu_{0}(t)+\Delta{\nu}_{p}+\Delta{\dot{\nu}}_{p} t + \Delta{\nu}_{d} e^{-t/\tau_{d}} \, ,
\label{glitch_nu}
\end{equation}
\begin{equation}
\dot{\nu}(t)= \dot{\nu}_{0}(t)+\Delta{\dot{\nu}}_{p}+\Delta{\dot{\nu}}_{d}e^{-t/\tau_{d}} \, ,
\label{glitch_nudot}
\end{equation}

where $\nu_{0}$ and $\dot{\nu}_{0}$ are pulse frequency and its derivative from per-glitch timing model, $\Delta{\nu}_{p}$ and $\Delta{\dot{\nu}}_{p}$ are the permanent changes of frequency and its derivative of the post-glitch relative to the pre-glitch values, $\Delta{\nu}_{d}$ and $\Delta{\dot{\nu}}_{d}$ represent the amplitude of the exponential decay on a timescale $\tau_{d}$.
\section{Analysis and results}
\label{res}
To investigate the long-term timing behaviors and the associated pulse profile changes of PSR B2035$+$36, observational data between Aug. 2002 and Aug. 2012 (MJD $52496\sim55899$) were extracted. We adopted astronomic ephemeris given by \citet{2004_Hobbs_pulsars} to obtain the best fitted timing parameters. The rotational parameters of PSR B2035$+$36 derived by fitting the timing model to pre- and post-glitch data are presented in Table \ref{rotation_parameters}. Uncertainties in the fitted parameters were taken to be twice the formal uncertainties obtained from \textsc{TEMPO2}.
\subsection{The glitch activity and spin-down rate change in PSR B2035$+$36}
\begin{figure}
\centering
\includegraphics[width=0.4\textwidth,angle=-90]{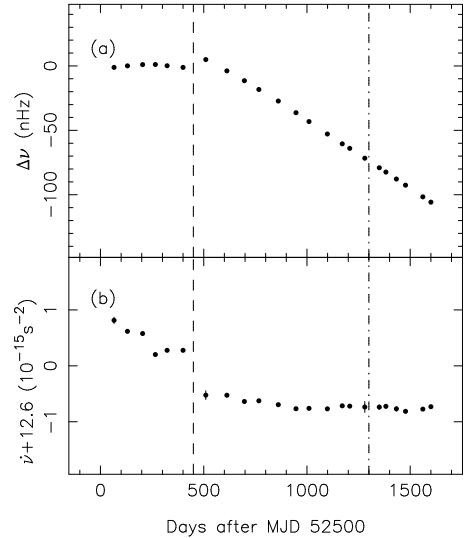}
\caption{Variations of $\nu$ and $\dot{\nu}$ of PSR B2035$+$36 by fitting $\nu$ and $\dot{\nu}$ for small sections of data. Each section contains about $15$ TOAs and repeats for $10$ ToAs, (a) variations of frequency $\Delta{\nu}$ relative to the pre-glitch solutions; (b) variations of the first frequency derivatives. The dashed line is the glitch epoch of MJD 52950, and the dot-dashed line is the epoch at MJD 53800.}
\label{2037+3621}
\end{figure}

Fig. \ref{2037+3621} shows the variations of frequency $\nu$ and its first derivative $\dot{\nu}$ during our data span, derived by independent fitting $\nu$ and $\dot{\nu}$ to short sections of data, with each section containing about $15$ TOAs and repeating for $10$ ToAs. The top panel shows that there was a small jump in $\nu$ resulting from glitch around MJD $52950$ (dashed line) and the bottom panel shows that the spin-down rate $\left|\dot{\nu}\right|$ changed significantly accompanied with the glitch. By fitting Equation \ref{glitch_nu}, we obtained the glitch parameters, presented in Table \ref{glitch_parameters}. It was noted that as there was no obvious exponential recovery process observed, we excluded the term of the exponential decay in this fitting. Unusually, the spin frequency of post-glitch decreased monotonically over more than $1000$ days, resulting that the observed spin frequency was much smaller than the pre-glitch. This was caused by a significant permanent increase in $\left|\dot{\nu}\right|$ ($\Delta{\dot{\nu}_{p}}\sim-0.84(3)\times 10^{-15} \, \rm s^{-2}$) along with a small glitch ($\Delta{\nu}_{p}\sim12.4(5) \, \rm nHz$). The corresponding glitch size and fractional change in spin-down rate were $\Delta{\nu}/\nu\sim 7.7(8)\times10^{-9}$ and $\Delta{\dot{\nu}}/\dot{\nu}\sim 0.067(8)$, respectively.

It's worth noting that $\left| \dot{\nu} \right|$ increased persistently over $800$ days after the glitch, which was opposite to the typical post-glitch behavior. The average $\left| \dot{\nu}\right|$ of post-glitch was about $9.6\%$ larger than pre-glitch. What's more interesting is that the glitch also induced significant change in the evolution of $\left|\dot{\nu}\right|$, in other words, the spin-down state. As shown in the bottom panel of Fig. \ref{2037+3621}, compared with the variation trend of pre-glitch, $\left|\dot{\nu}\right|$ of post-glitch evolved to a stabler state. Additionally, we notice that the long term timing behavior of post-glitch underwent a visible change around MJD $53800$ (dot-dashed line). To investigate the detailed rotational behavior of PSR B2035$+$36, timing solutions were obtained separately. Compared with the first period of MJD 52985$\sim$53794, $\left|\dot{\nu}\right|$ of the second period (MJD 53816$\sim$55899) was much stable with a smaller $\left|\ddot{\nu}\right|$.

\begin{table*}
\begin{center}
\caption{Timing parameters of PSR B2035$+$36.}
\label{rotation_parameters}
\begin{tabular}{lccc}
\hline\hline
Parameters & Pre-glitch & \multicolumn{2}{c}{Post-glitch} \\
\hline
Pulsar Name & \multicolumn{3}{c}{B2035$+$36(J2037$+$3621)} \\
R.A.(h:m:s)   & \multicolumn{3}{c}{$20:37:27.44(3)$}\\
Decl.($^{\circ}$ : $'$ : $''$)  & \multicolumn{3}{c}{$+36:21:24.1(3)$}\\

Pulse frequency, $\nu$ ($\, \rm s^{-1}$) & $1.61625000020(6)$ & $ 1.61624924998(3)$ & $1.616247561227(9)$\\
First derivative of pulse frequency, $\dot{\nu}$ ($\, \rm s^{-2}$) &  $-1.2037(5)\times 10^{-14}$ & $-1.3258(1)\times10^{-14}$ & $-1.32670(1)\times 10^{-14}$  \\
Second derivative of pulse frequency, $\ddot{\nu}$ ($\, \rm s^{-3}$) & $-2.3(2)\times10^{-23}$ & $-5.2(2)\times10^{-24}$ & $8.32(7)\times10^{-25}$ \\
Data Span (MJD) & $52496$ - $52907$ & $52985$ - $53794$ & $53816$ - $55899$ \\
Zero epoch for the timing solution (MJD) & $52701$ & $53389$ & $54858$ \\
Number of ToAs & $37$ & $58$ & $216$ \\
Rms timing residual ($\rm \mu s$) & $912$ & $1280$ &$1234$ \\

Time Scale & \multicolumn{3}{c}{TDB}\\
Solar system ephemeris model & \multicolumn{3}{c}{DE 421} \\
\hline
\end{tabular}
\end{center}
%Notes:a,\citep{2004_Hobbs_pulsars}.
\end{table*}

\begin{table}
\begin{center}
\caption{Glitch parameters of PSR B2035$+$36, which are produced by TEMPO2.}
\label{glitch_parameters}
\begin{tabular}{lc}
\hline\hline
Parameters & Value \\
\hline
Data Span (MJD) & $52705$ - $53360$ \\
Glitch Epoch(MJD) & $52950(40)$ \\
$\Delta{\nu}_{p}$ ($\rm s^{-1}$) & $12.4(5)\times10^{-9}$ \\
$\Delta{\nu}/{\nu}$ & $7.7(8)\times 10^{-9}$ \\
$\Delta{\dot{\nu}}_{p}$ ($\rm s^{-2}$) & $-0.84(3)\times10^{-15}$ \\
$\Delta{\dot{\nu}}/\dot{\nu}$ & $67(8)\times10^{-3}$ \\

\hline
\end{tabular}
\end{center}
\end{table}
\subsection{Mode changing}
\begin{figure}
\centering
\includegraphics[width=0.4\textwidth]{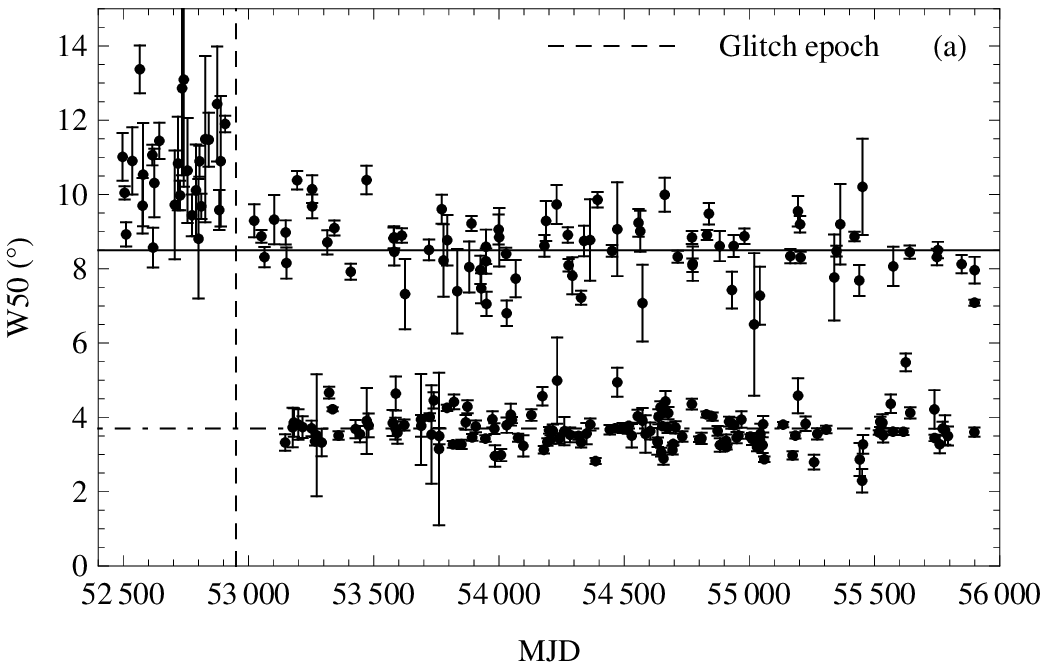}
\includegraphics[width=0.38\textwidth]{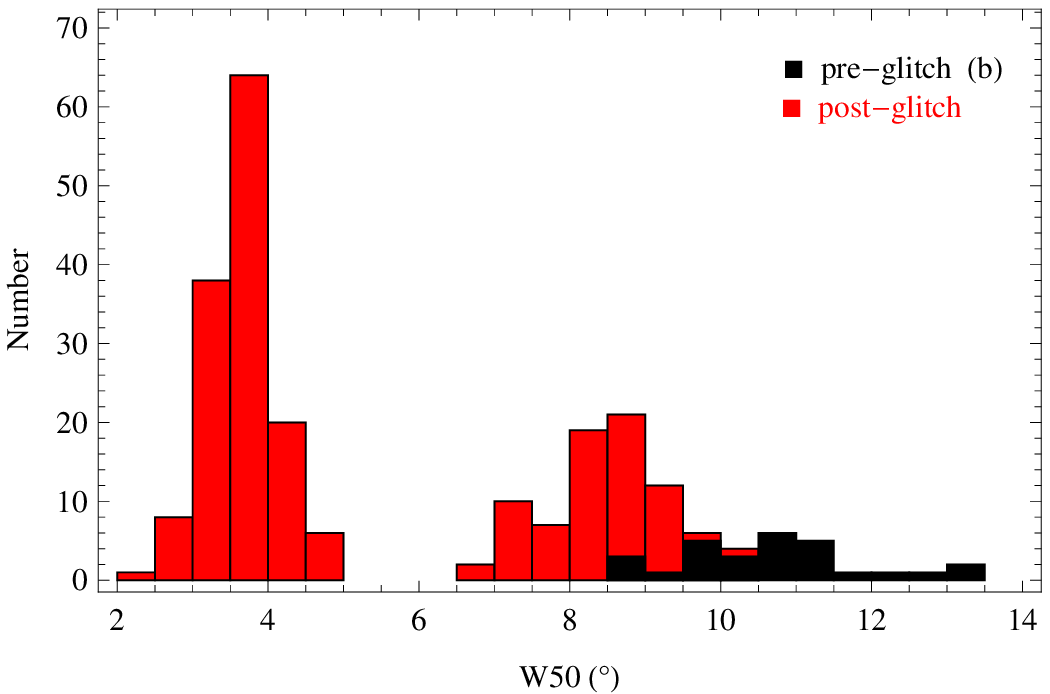}
\caption{Distributions of $W_{50}$ in unit of degree for PSR B2035$+$36, (a) distributions of $W_{50}$ as observation dates (MJD), the black solid line and dot-dashed line are the average values of $W_{50}$ ($8.5(7)^{\circ}$ and $3.7(2)^{\circ}$) of post-glitch; (b) the statistical distributions of $W_{50}$, the black and the red histograms are the pulse profiles of pre- and post-glitch, respectively.  }
\label{w50}
\end{figure}

\begin{figure}
\centering
\includegraphics[width=0.4\textwidth]{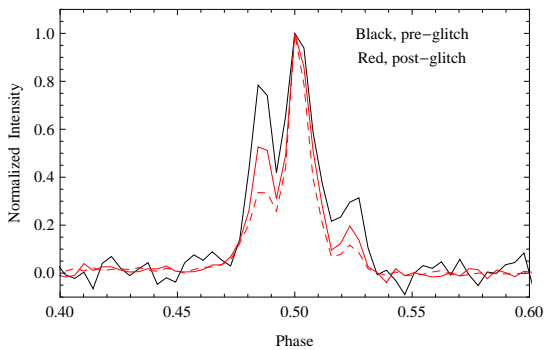}
\caption{The integrated normalized pulse profiles of different pulse profile modes of PSR B2035$+$36. The solid and the dashed lines indicate the wide and narrow pulse profiles, respectively. }
\label{pulse_profile}
\end{figure}

In addition to the spin-down state change, emission mode switching was also found in the post-glitch data. Distributions of the full width half maximum (FWHM and $W_{50}$ for short) of all integrated pulse profiles are given in Fig. \ref{w50}. The top sub-figure shows that the $W_{50}$ became narrower and began to switch between two states after the glitch (dashed line), the bottom sub-figure shows that there was obvious bimodal distribution of $W_{50}$ after the glitch. Visually, the number of the narrow pulse profiles was more than that of the wide ones. To analyze the pulse profile, we summed all pulse profiles according to different types to generated normalized integrated pulse profiles, which are presented in Fig. \ref{pulse_profile}. There was only one stable emission mode (pulse profile type) before the glitch (the black pulse profile). However, after the glitch, the pulsar switched between two emission modes with relatively narrow and wide pulse profiles (the solid and the dashed red pulse profiles), respectively. Besides, pulse profiles of post-glitch became narrower than pre-glitch. Basically, the pulse profile of PSR B2035$+$36 contains three components, and the middle component is dominant. The leading and the trailing components of post-glitch became weaker than pre-glitch, and because of intensity variation in the leading and the tail components, the pulsar switched between two emission modes.

Detailed parameters of pulse profiles are listed in Table \ref{profile_parameters}, where ``wide'' and ``narrow'' represent the pulse profiles with larger and smaller $W_{50}$. The average values of $W_{50}$ were $10.7(1.4)^{\circ}$ before the glitch, $8.5(7)^{\circ}$ and $3.7(3)^{\circ}$ (the solid line and dot-dashed line in (a) of Fig. \ref{w50}) after the glitch. The number of the narrow pulse profiles was about $1.5$ times of the wide ones, which indicates that the pulse profile of post-glitch was dominated by the narrow pulse profile. As discussed in the above subsection, there was a visible change in post-glitch timing behavior around MJD $53800$. We also analyzed the pulse profile independently. Generally, the bimodal distribution of $W_{50}$ and the mode changing behavior were consistent between MJD 52985-53794 and MJD 53816-55899. The number ratio of the narrow pulse profiles to the wide ones ($n_{\rm nar}/n_{\rm wid}$) increased from $\sim 1.32$  (MJD 52985-53794) to $\sim 1.83$ (MJD 53816-55899), which may indicate that the dominant trend of the narrow pulse profile became more and more obvious as pulsar evolving to a stabler rotation state.

\begin{figure}
\centering
\includegraphics[width=0.4\textwidth]{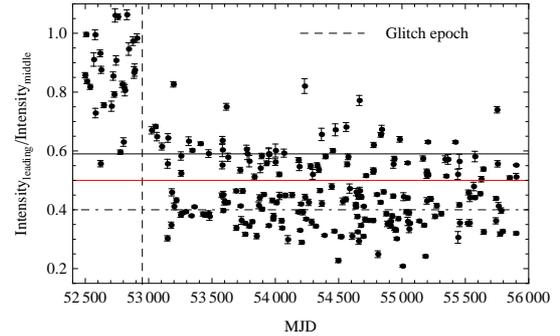}
\caption{Distributions of intensity ratio between the leading components and middle components for PSR B2035$+$36. The black solid line and the dot-dashed line are the average values of $0.57$ and $0.40$ of post-glitch. The red line is the boundary at $0.5$.}
\label{ratio}
\end{figure}

Fig. \ref{ratio} shows the distributions of the intensity ratios between the leading and middle components. Before the glitch, the intensity of leading components was comparable with the middle components with an average intensity ratio of $\sim 0.86(6)$. After the glitch, the intensity ratio was significantly smaller than its value of pre-glitch. Besides that, according to the intensity variations of the leading components, distributions of the ratios were also divided into two groups: the relatively large intensity ratios with mean value of $0.57(4) > 0.5$ and the relatively small intensity ratios with mean value of $0.40(3) <0.5$.

\begin{table*}
\begin{center}
\caption{Pulse profile parameters. The ``wide'' and the ``narrow'' mean the pulse profiles with larger and smaller values of $W_{50}$.}
\label{profile_parameters}
\begin{tabular}{lccccc}
\hline\hline
Parameters & Pre$-$glitch & \multicolumn{4}{c}{Post$-$glitch} \\
\hline
%Glitch Epoch (MJD) & \multicolumn{5}{c}{52950}\\
Data Span (MJD)  & $52496$ - $52907$ & \multicolumn{2}{c}{$52985$ - $53794$} & \multicolumn{2}{c}{$53816$ - $55899$}\\
           &               &  narrow & wide & narrow & wide \\

Mean FWHM $W_{50}$,($^{\circ}$) & $10.7(1.4)$ & $3.8(5)$ & $8.9(4)$ & $3.6(2)$ & $8.4(8)$ \\
Number of pulse profiles   & $28$ & $29$ & $22$ & $108$ & $59$ \\
Intensity ratio of leading and middle components & $0.86(6)$ & $0.40(3)$ & $0.62(5)$ & $0.39(3)$ & $0.57(4)$ \\
\hline

\end{tabular}
\end{center}
\end{table*}

\section{Discussions}
\label{dis}
This is the first detection for glitch in PSR B2035$+$36 since it was discovered in 1985 \citep{1985_dewey}. It is a relatively old pulsar but undergoes a small glitch, the relative change in spin-down rate of $\Delta{\dot{\nu}}/\dot{\nu}\sim 0.067(8)$ is relatively large compared with other small glitches \citep{2010_Yuan_glitch,2011_Espionza_glitch,2013_Yu_glitch,2017_Fuentes_glitch_actibity}. It is possible that a recovery process was missed because of the large data gap around the glitch, and the decay time should be $\leq80$ days. The small glitch could be understood either by crust rearrangement in the crust-quake model or by a sudden momentum transfer in the superfluid scenario \citep{1991_Ruderman_star_quake,1984_Alpar_vortex_creep}.

Unusually, the spin-down rate increased persistently over $800$ days after the glitch, which was similar with the post-glitch behavior of PSR J1718$-$3718 but opposite to the typical post-glitch behavior \citep{2011_Manchester_1718}. This abnormal evolution trend of $\left|\dot{\nu}\right|$ after the glitch could not be explained by standard glitch models. We thus assume that there should be change in external braking torque accompanied with the glitch activity, and a gradually increasing braking torque, for example, a variable out-flowing particle density in the magnetosphere \citep{2015_Kou&Tong}, will lead to the persistent increasing spin-down rate.

In our work, the average increase of $\left|\dot{\nu}\right|$ is $9.6\%$ \footnote{\citet{2010_Lyne_magnetosphere} presented $13.28\%$ increase in $\left|\dot{\nu}\right|$ by measuring
the peak-to-peak values of $\dot{\nu}$ for PSR B2035+36. A $13.8(6)\%$ increase in $\left|\dot{\nu}\right|$ was given by measuring the maximum and minimum values of $\dot{\nu}$ using our data. We would like to adopt the average value of $9.6\%$ increase in $\left|\dot{\nu}\right|$ for further discussions.}. An additional braking torque may be induced by glitch activity, leading to the relatively large permanent increase in $\left|\dot{\nu}\right|$. In the wind braking scenario, the spin-down behavior is affected by the out flowing particle density $\frac{\dot{\Omega}^{'}}{\dot\Omega}=\frac{\eta(\kappa^{'})}{\eta(\kappa)}$, where $\Omega=2\pi\nu$ is the angular velocity, $\kappa$ means that the accelerated particle density is $\kappa$ times of Goldreich-Julian charge density ($\rho_{\rm e}=\kappa \rho_{\rm GJ}$), and $\eta=\sin^2\alpha+4.96 \times 10^{2} \kappa B_{12}^{-8/7} \Omega^{-15/7} \cos^2 \alpha$ in the vacuum gap model \citep{2001_xu_wind}. Glitch may induce the fluctuation of magnetosphere, and hence the process of particle acceleration and radiation \citep{2015_Kou&Tong}. To explain the $9.6\%$ increase in spin-down rate, $22\%$ increase in the particle density is needed in the vacuum gap case \citep{2016_Kou_variable_timing_0540}. However, in the MHD (short for magnetohydrodynamic) simulation, the relative change of $\left|\dot{\nu}\right|$ can be linked to the change of inclination angle $\Delta{\dot{\nu}}/\dot{\nu}=\sin{2\alpha}\Delta{\alpha}/(1+\sin^{2}{\alpha})$ \citep{2006_Spitkovsky_MHD,2017_Zhao_2021}. Glitch may change the magnetic field structure and hence the inclination angle \citep{2016_Ng}. Corresponding to the $9.6\%$ increase in spin-down rate, the expected change in inclination angle is $\Delta{\alpha}\sim 8^{\circ}$ if the inclination angle of $45^{\circ}$ is assumed for this pulsar, and the fluctuation of inclination angle will result in the change of effective emission geometry, hence the observed pulse profile variation.

Pulse profiles of PSR B2035$+$36 are integrated over $4\sim16 \, \rm mins$, which may contain both two kinds of pulse profile (wide and narrow). The wide and narrow integrated pulse profiles may be performances of pulse profiles dominated by wide or narrow individual pulses. The time scale of two emission modes and the interval between their occurrences need longer single pulse observations using larger radio telescope. Future observations will enable more accurate estimations. The averaged pulse profiles show that the pulsar switches between different emission modes with the intensity variation in leading and tailing components, which indicate that the emission state switches between different magnetospheric states. The observation may be a manifestations of a nonuniform distribution of particle or shrinking and expanding of the emission zone \citep{2010_Timokhin}. In addition to the short term switching between two emission modes, it seems that the pulse is gradually dominated by the relatively narrow pulse profile mode over the long term. This may be a manifestation of the magnetosphere gradually evolving to a stabler state.

The observations of profile changes associated with the variable timing behavior in several pulsars indicate that their rotations are correlated with their emissions \citep{2010_Lyne_magnetosphere}. It is not clear for long what induced the magnetospheric activity. Glitch may be a trigger mechanism. PSR J1119$-$6127 is a young, isolated radio pulsar with high surface magnetic field. There were RRAT-like and intermittent-like emissions observed following an unusual glitch activity \citep{2011_Weltevrede_1119}. PSR J0742$-$2822 is a normal isolated radio pulsar, a strong correlation between the spin-down rate and pulse shape was observed after a glitch \citep{2013_keith-0742}. PSR B2035$+$36 is the third radio pulsar to show pulse profile change and spin-down rate change directly accompanied with glitch activity.  Besides, PSR J2021$+$4026 stayed at a high spin-down rate and low gamma ray emission state for about $3$ years after the glitch \citep{2017_Zhao_2021}. Glitch sizes of these four pulsars range from $7\times 10^{-9}$ to $5\times10^{-6}$, with their characteristic age spanning from $1.6 \, \rm kyr$ to $2000 \, \rm kyr$. It seems that the ``glitch-induced'' magnetospheric behavior occurred in both relatively young pulsars and old pulsars, and has no obvious relation to the glitch size. More such samples will provide us a hint to understand the emission mechanism of pulsars.
\section{Conclusions}
\label{con}
Glitch is usually believed as the internal origin. Furthermore, the spin-down and emission behavior is thought to be driven from the external braking torque and magnetospheric radiation. For PSR B2035$+$36, a permanent increase in the spin-down rate and pulse profile change were found accompanied with a glitch activity occurred at MJD $52950$. Besides, mode change was also observed in the post-glitch data, and the relatively narrow pulse mode gradually became dominant. Observations of PSR B2035$+$36 present a new direct observational evidence about the connection between magnetosphere behavior and glitch activity, which suggest one possibility that the magnetospheric fluctuation may be triggered by glitch event.

\section*{Acknowledgments}
The authors would like to thank K. S. Cheng for discussions and Y. Zhang for language editing. This work is based on observations made with the Urumqi Nanshan 25-m Telescope, and supported by the National Basic Research Program of China grants 973 Programs (2015CB857100), the National Key Research and Development Program of China (No. 2016YFA0400800, 2017YFA0402602). J.P.Y. is supported by the Strategic Priority Research Program of the Chinese Academy of Sciences, Grant No. XD23010200. W.M.Y. acknowledges support from West light Foundation of CAS (No. XBBS201422) and National Natural Science Foundation of China (Nos. U1631106, U1731238).

%%%%%%%%%%%%%%%%%%%%%%%%%%%%%%%%%%%%%%%%%%%%%%%%%%



%\bibliographystyle{mnras}
%\begin{thebibliography}{99}
%\bibliography{2037}
%\end{thebibliography}

%%%%%%%%%%%%%%%%%%%%%%%%%%%%%%%%%%%%%%%%%%%%%%%%%%

%%%%%%%%%%%%%%%%% APPENDICES %%%%%%%%%%%%%%%%%%%%%

%\appendix



%%%%%%%%%%%%%%%%%%%%%%%%%%%%%%%%%%%%%%%%%%%%%%%%%%

% Don't change these lines
\bsp	% typesetting comment
\label{lastpage}
\end{document}